\documentclass[aps,twocolumn,secnumarabic,balancelastpage,amsmath,amssymb,nofootinbib,hyperref=pdftex]{revtex4}

\usepackage{chapterbib}    
\usepackage{color}         
\usepackage{graphics}      
\usepackage[pdftex]{graphicx}      
\usepackage{longtable}     
\usepackage{epsf}          
\usepackage{bm}  
\usepackage{float}
\usepackage{verbatim}
\usepackage{svg}
\usepackage{soul}          
\usepackage[colorlinks=true]{hyperref}  

\usepackage{amssymb}
\usepackage{amsmath}

\begin{document}
\title{Optically Pumped AlGaN Double Heterostructure Deep-UV Laser by Molecular Beam Homoepitaxy: Mirror Imperfections and Cavity Loss}
\author {Len van Deurzen$^{1,}$}
\email{lhv9@cornell.edu}
\author{Ryan Page$^{2}$}
\author{Vladimir Protasenko$^{3}$}
\author{Huili (Grace) Xing$^{2,3}$}
\author{Debdeep Jena$^{1,2,3}$}

\affiliation{$^{1}$Department of Applied and Engineering Physics, Cornell University, Ithaca, USA}
\affiliation{$^{2}$Department of Materials Science and Engineering, Cornell University, Ithaca, USA}
\affiliation{$^{3}$Department of Electrical and Computer Engineering, Cornell University, Ithaca, USA}

\begin{abstract}
We demonstrate the first optically pumped sub-300 nm UV laser structures grown by plasma-assisted molecular beam epitaxy on single-crystal bulk AlN. The edge-emitting laser structures fabricated with the AlN/AlGaN heterostructures exhibit multi-mode emission with peak gain at ~284 nm. Having the goal of electrically injected, continuous wave deep-UV AlGaN laser diodes in mind, with its intrinsic material challenges of achieving sufficient optical gain, the optical cavity loss of a laser diode should be minimized. We derive an expression to quantify the effect of mirror imperfections, including slant and surface roughness on the optical mirror loss of a Fabry-P\'erot cavity. It is found that the optical imperfection loss is a superlinear function of the RMS roughness and slant angle of the facets, and also scales as the inverse wavelength squared of the principal lasing mode. This highlights the importance of device processing optimization as Fabry-P\'erot cavities couple to lower wavelengths. 
\end{abstract}

\maketitle


High-power, semiconductor UV-B (280-315 nm) and UV-C (100-280 nm) emitters are desirable as they are an energy-efficient and compact tool with diverse applications. These include pathogen detection and sterilization \cite{welch_far-uvc_2018, cheng_inactivation_2020}, water purification \cite{mezzanotte_wastewater_2007}, gas sensing \cite{mehnke_gas_2017} photolithography \cite{allcock_ultraviolet_2006}, and quantum computing and metrology \cite{ozawa_single_2017, taylor_quantum_2016}, to name a few. The realization of UV-C laser diodes (LDs) based on the AlGaN semiconductor material system has proven challenging, as these wide-bandgap semiconductors exhibit low carrier mobilities, very large dopant activation energies, resulting in low free carrier concentrations, and asymmetries between electron and hole transport \cite{park_fundamental_2018, amano_2020_2020}. This makes it a challenge to achieve population inversion and sufficient gain by electrical injection. Since lasing occurs when the optical gain of the laser diode equals its optical losses, the minimization of the optical losses is of paramount importance. 

Optically pumped AlGaN lasers, as well as pulsed electrically pumped laser diodes have recently been achieved by metal organic chemical vapor deposition (MOCVD).  \cite{xie_lasing_2013,zhang_2718_2019, sato_room-temperature_2020}. Optically pumped lasing has also been achieved in AlGaN multiple quantum wells grown by plasma-assisted molecular beam epitaxy (PA-MBE) on sapphire \cite{jmerik_optically_2010, jmerik_plasma-assisted_2013}. However, heteroepitaxy on sapphire results in a large threading dislocation density of $\gtrapprox 10^{8}$/cm$^{2}$, leading to a large optical intrinsic loss and reduced optical gain due to the scattering of the optical cavity modes and free carriers \cite{hasegawa_dislocation-related_2007, jena_dislocation_2000}. With the recent development of molecular beam homoepitaxy on bulk AlN substrates, the optical intrinsic loss can be kept low by reducing the threading dislocation density $\approx$ millionfold, as well as reduce the point defect density compared to growth on foreign substrates such as SiC and sapphire and templates  \cite{cho_molecular_2020,lee_mbe_2021}. MBE also offers certain advantages over MOCVD. For instance, MBE-grown thin films show a low O, H and C impurity background, and atomically sharp heterostructure interfaces can be grown. This is useful for the growth of tunnel junctions, short-period superlattices (SPSLs), or monolayer thick multiple quantum wells \cite{islam_mbe-grown_2017}. Moreover, there are no memory- or activation-effects present for Mg doped p-GaN, allowing for the growth of buried, conductive p-type layers. This enables inverted diode structures, where spontaneous polarization fields stemming from the heterostructure interfaces are favorably aligned with the applied field to enhance carrier injection and reduce carrier overflow. This has recently been achieved in tunnel junction based (n-p-i-n) AlN/GaN/InGaN light-emitting diodes \cite{bharadwaj_enhanced_2020, lee_light-emitting_2020}.

In the first section of this work, we demonstrate the first optically pumped sub-300 nm AlGaN laser grown by molecular beam epitaxy on single crystal AlN. The double heterostructure (DH) Fabry-P\'erot laser bars exhibit multi-mode emission with peak gain at $\approx$ 284 nm.

Besides reducing the optical intrinsic loss by growth on bulk AlN, it is important to limit the optical cavity loss. In Fabry-P\'erot laser diodes, the facets often exhibit imperfections such as slanting and roughness. These imperfections cause a phase shift of the specularly reflected cavity modes, resulting in optical cavity loss. In the second section of this paper, we derive an expression showing that this optical imperfection loss is a superlinear function of the slant angle and facet RMS roughness, and that their prominence is more severe for decreasing wavelength, scaling as the inverse modal wavelength squared.  To this purpose, combined dry etching and cleaving with a TMAH-based wet etching technique is applied to form smooth, vertical mirrors \cite{yasue_dependence_2019}. The avoidance of significant mirror imperfection losses and the demonstration of the optically pumped laser are crucial for the realization of the more desirable electrically injected, continuous wave, UV-B and UV-C laser diodes by MBE.

\begin{figure*}[t]
\includegraphics[width=18cm]{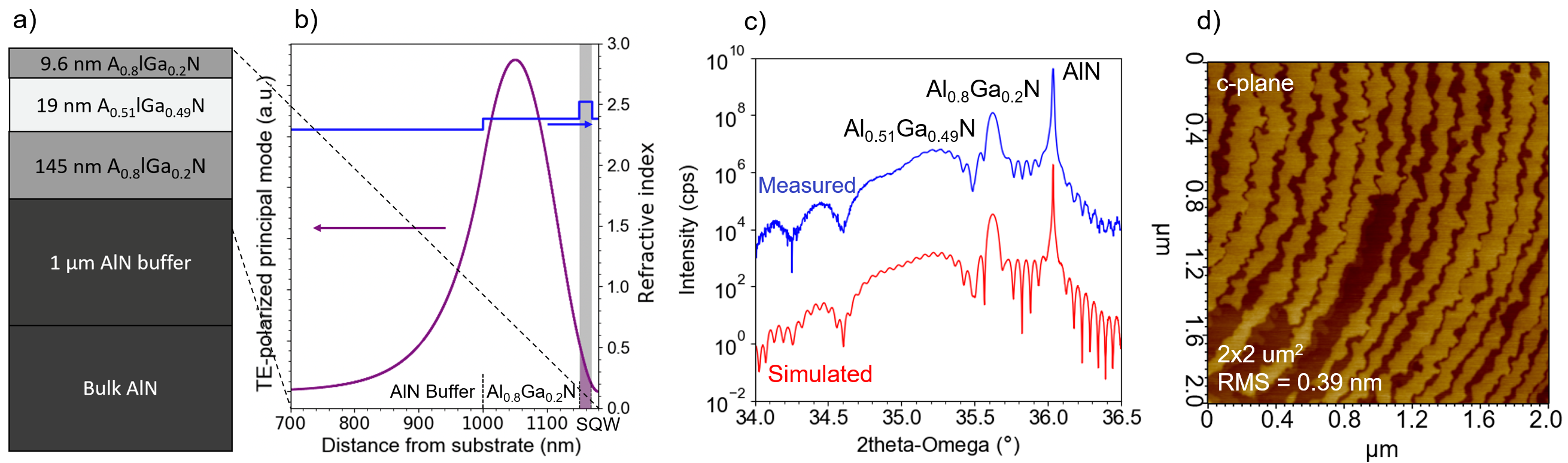}
\caption{(a) Schematic of the double heterostructure AlGaN laser with (b) the simulated principal TE optical mode and refractive index. (c) Measured (blue) and simulated (red) 2$\theta$-$\omega$ x-ray diffraction scans of the quantum heterostructure in (a) around the [0002] wurtzite symmetry axis. (d) Atomic force microscopy image of the c-plane surface with distinct atomic steps.   \label{fig:Figure_1}}
\end{figure*}


The double heterostructure lasers reported here are grown by PA-MBE on Al-polar, c-plane single crystal AlN substrates with a dislocation density on the order of $10^4$ cm$^{-2}$. The epistructural layer schematic is shown in Figure \ref{fig:Figure_1}(a). Prior to introduction to the MBE chamber, the substrates are cleaned with acids to partially remove surface oxides and contaminants, following a procedure reported previously \cite{cho_molecular_2020}. Immediately before growth, the substrates undergo a second cleaning process consisting of repeated cycles of aluminum deposition and thermal desorption to further remove native oxide layers and enable dislocation-free nucleation. For all samples, a high substrate temperature (T$_{sub}\approx1000^{\circ}$C) AlN buffer layer is grown with a thickness of 1 $\mu$m to isolate the active region of the device from the remaining impurities present at the substrate interface, which can contribute to optical loss. Following the buffer layer growth, all excess Al is thermally desorbed and the substrate temperature is lowered to $\approx 800^{\circ}$C for the growth of the subsequent AlGaN layers. A bottom waveguide is then grown, targeted at a thickness of 150 nm to remove the optical optical field intensity of the principal lasing mode from the AlN-Al$_{0.8}$Ga$_{0.2}$N interface, to limit the optical loss from surface-riding impurities formed around that interface \cite{chaudhuri_high-conductivity_2021}. Then, a 20 nm thick quantum well gain layer is grown aimed at an aluminum composition of 55$\%$, followed by a 10 nm Al$_{0.8}$Ga$_{0.2}$N cap layer. Excess metal is thermally desorbed from the crystal surface at each heterointerface to ensure sharp interfaces and abrupt compositional change. 

The cap layer thickness is chosen to maximize the product of the optical confinement factor, $\Gamma$, and the optical gain, $g$. For the quantum well composition of Al$_{0.55}$Ga$_{0.45}$N, the lasing optical modes are transverse electric (TE) polarized, meaning that for increasing cap layer thickness, $\Gamma$ increases as well \cite{zhang_effect_2010}. On the other hand, due to generation and recombination of electron-hole pairs in the Al$_{0.8}$Ga$_{0.2}$N cap layer upon optical pumping from the surface, the gain of the double heterostructure laser exponentially decreases as a function the cap layer thickness for a given pump intensity. These competing effects are simulated using an optical AlGaN heterostructure solver, SiLENSE. The resulting product $\Gamma g$ is found to be maximized at a cap layer thickness of 10 nm when pumping the double heterostructure from the topside with a 193 nm ArF excimer laser.  Such an ArF laser is employed as the source of population inversion for the AlGaN double heterostructure laser reported here. A schematic of the double heterostructure laser, and the accompanying principal TE polarized optical mode and the 2$\theta$-$\omega$ x-ray diffraction (XRD) scan and simulation around the [0002] wurtzite axis are shown in Figure \ref{fig:Figure_1}(a)-(c). The simulated XRD plot in Figure \ref{fig:Figure_1}(c) suggests the thicknesses and composition are close to the target and are given in Figure \ref{fig:Figure_1}(a). Figure~\ref{fig:Figure_1}(d) shows an atomic force microscopy image of the c-plane surface of the grown double heterostructure laser. Atomic steps due to the miscut angle of the substrate surface are visible. Along with the very low RMS roughness of 0.39 nm, they indicate that the heterostructure was successfully grown in the desired two-dimensional growth mode. 

\begin{figure*}[t]
\includegraphics[width=18cm]{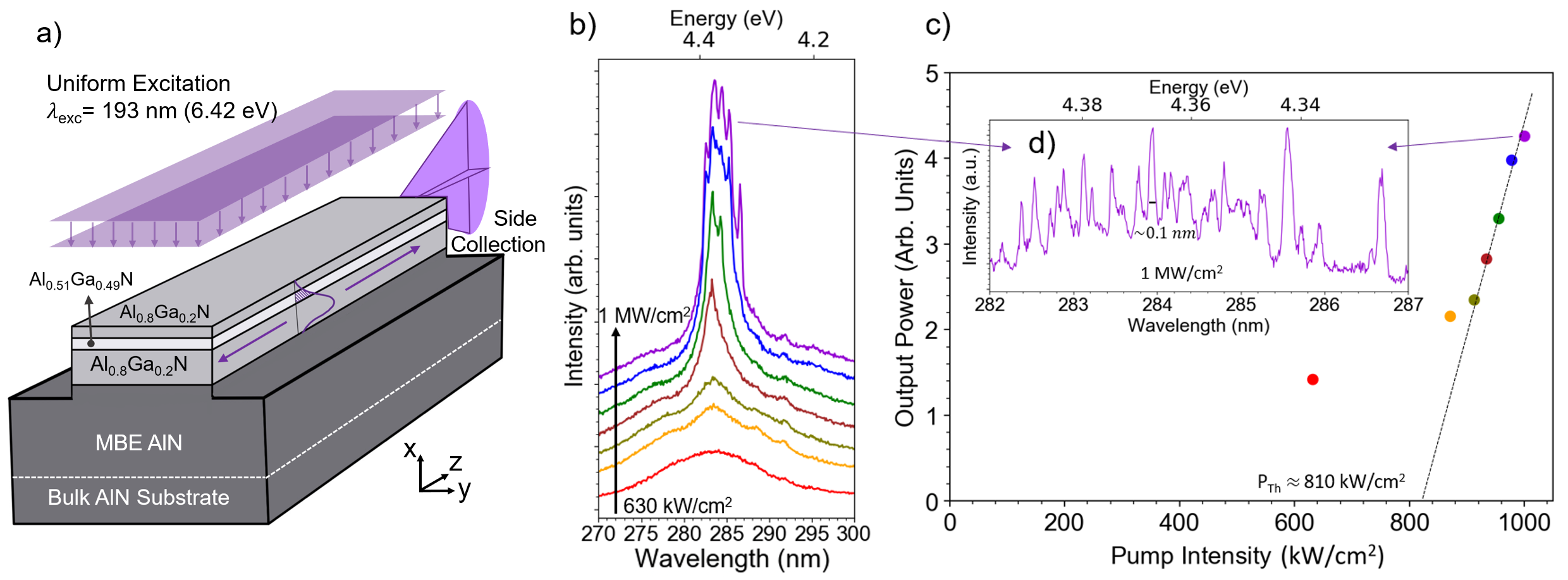}
\caption{(a)  Schematic of an optically pumped double heterostructure AlGAN laser bar. (b) Low resolution emission spectra at different excitation powers. (c) Output power vs. input power, showing lasing threshold at $\approx$ 900 kW/cm$^{2}$. (d) High resolution multi-mode lasing spectrum at an excitation intensity of $\approx$ 1 MW/cm$^{2}$. \label{fig:Figure_2}}
\end{figure*}

The epistructure in Figure \ref{fig:Figure_1}(a) was fabricated into both etched facet emitters as well as cleaved facet emitters, as shown in the schematic in Figure \ref{fig:Figure_2}(a). The sidewalls and facets of the etched facet emitters are formed by inductively coupled reactive-ion etching (ICP-RIE) followed by a tetramethylammonium hydroxide (TMAH) wet etch at 85~$^{\circ}$C for 10 minutes. The cleaved facet emitters are cleaved after ICP-RIE, followed by a similar TMAH wet etch. The TMAH etch converts the slanted facets and sidewalls resulting from the dry etch or roughened $m$- and $a$-plane surfaces to vertical, smoother walls \cite{yasue_dependence_2019}. The avoidance of these mirror imperfections is crucial to achieve significant optical feedback from the facets, as described in the next paragraphs.

An ArF excimer laser emitting at 193 nm (6.42 eV) is used as the optical pump for the AlGaN laser bars and is pulsed at a width of 10~ns at 2 mJ per pulse with a repetition frequency of 100 Hz. The excitation pattern is focused into a pencil shape by cylindrical lenses for uniform excitation of the top of the DH laser, and the intensity is varied through the use of deep-ultraviolet optical filters. The emission is collected from the facet of the DH laser and its spectrum is measured with a 500 mm focal length spectrometer with diffraction grating of 2400 lines per millimeter with 240 nm blaze wavelength. This is illustrated in the schematic in Figure \ref{fig:Figure_2}(a).

Figure \ref{fig:Figure_2}(b) shows the low resolution photoluminescence spectra for a $\approx$ 1 mm long cleaved facet laser for various excitation intensities. As the pump intensity is increased, the spectral FWHM decreases from $\approx$ 15 nm at 630~kW/cm$^{2}$ to $\approx$ 5 nm at 900~kW/cm$^{2}$ due to the onset of amplified spontaneous emission. From the output versus input power plot in Figure \ref{fig:Figure_2}(c), the threshold of light amplification by stimulated emission occurs at $\approx$~810~kW/cm$^{2}$. At 1 MW/cm$^{2}$, this is accompanied with a broad gain spectrum with peak gain at $\approx$~284~nm and multi-mode emission is realized, as shown in Figure~\ref{fig:Figure_2}(d). The resolution of the spectrometer is limited to $\approx$ 0.01 nm. However, for a laser bar of longitudinal length $L$ and effective refractive index $n_{eff}$ , emitting at wavelength $\lambda$, the modal spacing $\Delta \lambda$ is approximately equal to $ \lambda^{2}/2Ln_{eff}$ \cite{Sze_physics_2008}. For $L$ $\approx$ 1 mm and $n_{eff} \approx 3$, $\Delta \lambda$ $\approx$ 0.013 nm \cite{guo_impact_2020}. Since $\Delta \lambda$ is close to the spectrometer resolution, and some of the modes are suppressed, not all are resolved. The FWHM of the resolved modes is $\approx$ 0.1 nm. It must be noted that a there is a large $\approx$~2.06 eV of heat generated by phonons due to hot electron-hole pair generation by each absorbed 193 nm ($\approx$ 6.42 eV) photon. Therefore, the lasing threshold pump intensity can be significantly decreased by increasing the quantum well and cladding layer aluminum compositions (increasing bandgap) or increasing (decreasing) the pump wavelength (energy). 

\begin{figure*}[t]
\includegraphics[width=18cm]{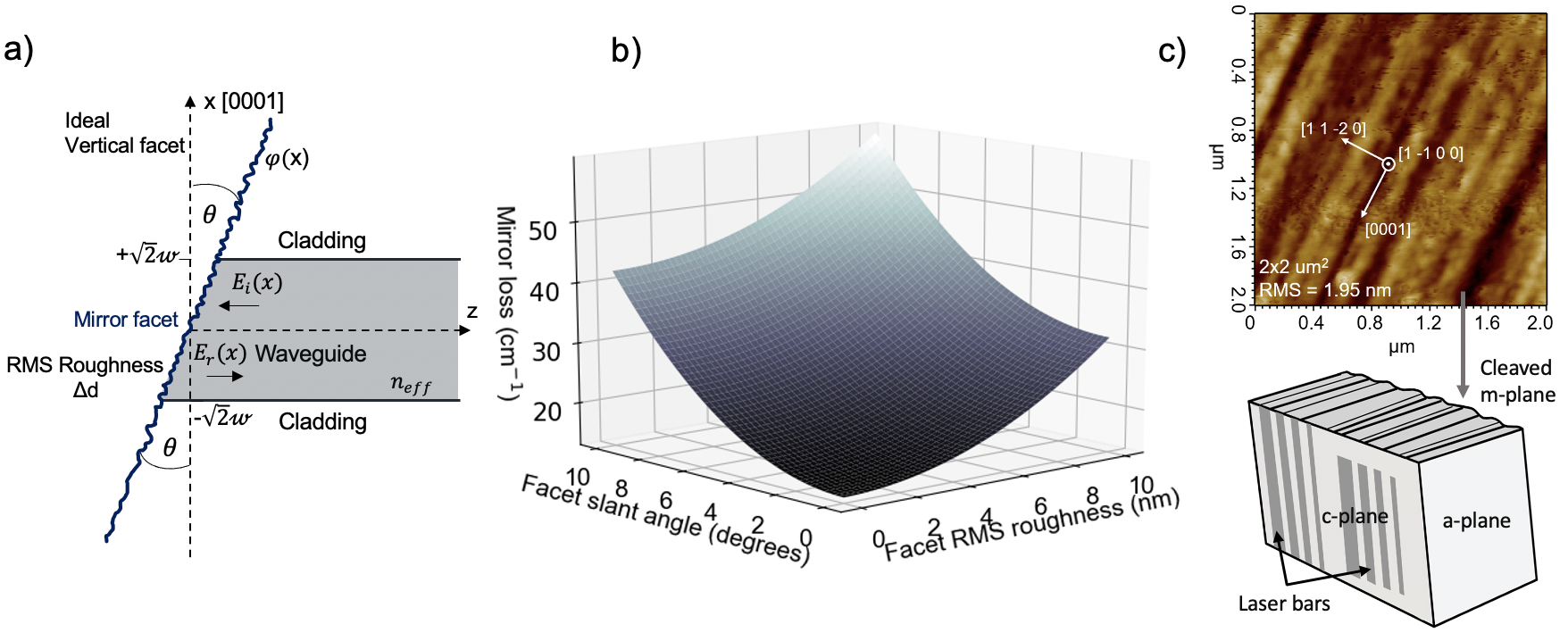}
\caption{(a) Schematic depicting the mirror imperfections of slant and facet roughness for a Fabry-P\'erot waveguide. (b) Mirror loss as a function of facet slant angle and surface RMS roughness, for a laser bar of length L = 1 mm, $\lambda$ = 280 nm, $n_{eff}$ = 3, and $w$ = 50 nm. (c) 2x2 $\mu$m$^{2}$ atomic force microscopy image of an m-plane facet after cleaving and TMAH etching, with an RMS roughness of 1.95 nm corresponding to a theoretical imperfection loss of $\approx$ 1.4~cm$^{-2}$. \label{fig:Figure_3}}
\end{figure*}

An increasingly important aspect of cavity optics at short wavelength photonic devices is the sensitivity to boundary imperfections, which can be a specially severe bottleneck for deep-UV laser diodes. In order to quantify the importance of laser bar processing optimization, a model for the effect of mirror imperfections on the optical cavity loss of a Fabry-P\'erot laser structure is now discussed. To achieve light amplification by stimulated emission, the lasing equation must be satisfied: 
\begin{equation}
\Gamma g = \alpha_{intrinsic}+ \alpha_{cavity},
\label{Laser_Equation}
\end{equation} 
where $\Gamma$ is the confinement factor, $g$ the optical gain, $\alpha_{intrinsic}$ the optical intrinsic (material) loss and $\alpha_{cavity}$ the optical cavity loss. $\alpha_{cavity}$ is the sum of the mirror loss, $\alpha_{m}$,  and ridge loss, $\alpha_{R} $, where the latter is neglected in this discussion.  For a Fabry-P\'erot cavity with two identical mirrors and length $L$, the mirror loss is given by: 

\begin{equation}
    \alpha_{m} = \frac{1}{L} \ln\frac{1}{R}.
\label{IdealMirror_Equation}
\end{equation}
Often, the mirrors are treated as idealized facets defined by a single semiconductor-air interface and reflectance $R_{0}$=$(n-1)^{2}/(n+1)^{2}$. However, in practice, the facets often exhibit imperfections such as tilt due to inductively coupled plasma reactive ion etching (ICP-RIE) or cleaving and roughening of the semiconductor-air interface due process inhomogeneities, anisotropic etch rates, or cleaving. 

The effects of these imperfections on mirror reflectivity have been discussed seperately by Iga et al. and Stocker et al., respectively \cite{iga_mode_1981,stocker_facet_1998}. Both of the imperfections cause a phase shift of the specularly reflected longitudinal modes of the laser, reducing the net reflectance of the mirrors. The combined effect of facet tilt and roughness is discussed now. This scenario is illustrated in Figure~\ref{fig:Figure_3}(a). First, consider the phase shift due to a facet that is tilted by an angle $\theta$ with respect to the vertical ($x$-axis). At $z=0$, the TE polarized optical mode, which is the dominant mode for AlGaN quantum wells with a quantum well Aluminum composition below $\approx 70\%$, can be approximated as a Gaussian beam with waist radius $ \sqrt2 w$ at the mirror interface, where $w$ is the beam waist \cite{zhang_effect_2010}. The incident, transverse principal field at the interface can then be written as:
\begin{equation}
E_{i}(x) \simeq E_{0} \left(\pi w^{2}\right)^{-1 / 4} e^{-\frac{1}{2}\left(\frac{x}{w}\right)^{2}},
\label{H00_Equation}
\end{equation}
where the pre-exponential term is the electric field amplitude, and $E_{i}(x, z=0) \approx E_{i}(x,z=\tan(\theta)x)$. The phase shift $\phi_{m}$ by the longitudinal mode reflected at each point along the mirror is equal to
\begin{equation}
\phi_{m} = 4 \pi n_{eff} x \tan(\theta) / \lambda,
\label{Phi_m}
\end{equation}
where $n_{eff}$ is the effective refractive index corresponding to the waveguide and $\lambda$ is the free-space wavelength corresponding to the longitudinal lasing mode.
Secondly, if the surface of the slanted facet is roughened, following a Gaussian distribution, then the RMS phase broadening $\Delta \phi_{RMS}$ can be found by regarding each point along the facet as a Huygens point source:
\begin{equation}
    \Delta \phi_{RMS} = 4 \pi n_{eff} \cos(\theta) \Delta d / \lambda,
\label{Roughness_phase}
\end{equation}
where $\Delta d$ is the RMS roughness of the facet surface. Then, the distribution, $P(\phi_{d})$, of the reflected beam having a phase shift $\phi_{d}$ due to the facet roughness, is given by:
\begin{equation}
P(\phi_{d})=\frac{e^{-\phi_{d}^{2} / (2 \Delta \phi_{RMS}^{2})}}{{\sqrt{2 \pi} \Delta \phi_{RMS}}}.
\label{Distribution_Phi_d}
\end{equation}

Finally, assuming a semiconductor-air reflectivity, $r_{0} = (n_{eff}-1)/(n_{eff}+1)$, and that the facet roughness occurs at a length scale $<< w$, we can calculate the net reflectivity, $r$, from the mirror imperfections and incident beam $E_{i}(x)$:

\begin{equation}
\begin{split}
r & \simeq r_{0}   \int_{-\infty}^{\infty}\int_{-\infty}^{\infty} P(\phi_{d}) E_{i}(x) e^{i(\phi_{d}+\phi_{m})}  d \phi_{d} d \phi_{m}  / N \\
& = r_{0} e^{-\frac{(4n_{eff} \pi)^{2}\left((\cos (\theta) \Delta d)^{2}+(w \tan (\theta))^{2}\right)}{2 \lambda^{2}}} ,
\end{split}
\label{ReflectedAmplitude_Integral}
\end{equation}

where $N = \int_{-\infty}^{\infty} E_{i}(x) d \phi_{m}$. Using $R = r^{2}$ and Equation \ref{IdealMirror_Equation}, the final mirror loss of a laser with two identical mirrors with the described imperfections can be written as:

\begin{equation}
\begin{split}
\alpha_{m} = & \frac{(4 \pi)^{2}}{(\lambda / n_{eff})^{2} L} \left[(\Delta d \cos(\theta))^{2} + (w \tan (\theta))^{2} \right] \\ & +\frac{2}{L} \ln \left(\frac{n_{eff}+1}{n_{eff}-1}\right).
\end{split}
\label{Imperfection_Loss}
\end{equation}

This implies that the mirror imperfection loss scales as the inverse wavelength squared of the longitudinal lasing mode.
For example, consider an AlGaN deep-UV cavity with $\lambda$~=~250 nm and an InGaN blue cavity with $\lambda$~=~450 nm, with $n_{eff, InGaN}$/$n_{eff, AlGaN} = 1.1$ \cite{, leung_refractive_1998, muth_absorption_1999}. The AlGaN cavity exhibits a mirror imperfection loss term $\approx$ 2.5 times as large as an InGaN blue cavity for the same degree of tilt and RMS roughness. For a laser structure similar to the DH discussed here, Figure~\ref{fig:Figure_3}(b) shows a plot of mirror loss as a function of facet slant angle and surface RMS roughness, for $L$ = 1~mm, $\lambda$ = 280~nm, $n_{eff}$ = 3, and $w$ = 50~nm. For example, such a laser bar with $\theta =5.5^{\circ}$ and  $\Delta d =5.5$~nm corresponds to an approximate doubling of the mirror loss compared to the idealized case. It must be noted that facet slanting between the $z$- and $y$-axis can also occur, which roots a more severe optical mirror imperfection loss since the beam waist is larger in the $y$ direction. However, since this slanting is caused by misaligned cleaving, it is easier to avoid.
For the processed laser bars, facet RMS roughness as low as 1.95 nm was achieved with near-vertical facets, with an atomic force microscopy image of the facet shown in Figure~\ref{fig:Figure_3}(c). This value corresponds to a theoretical imperfection loss of merely $\approx$ 1.4~cm$^{-1}$. However, the threshold pump intensity is still fairly high compared to those reported by MOCVD \cite{kirste_6_2018}. 

Certain straightforward design and experimental optimizations can be made to reduce the threshold pump intensity. First of all, the active region can be optimized to enhance the gain, such as in a multiple quantum well design \cite{witzigmann_calculation_2020}. Or, by reducing the difference in Al composition between the quantum well and barriers, the polarization fields are screened at a lower pump intensity. The intrinsic optical loss can be reduced by removing the optical mode from the AlN-AlGaN interface, which can be achieved by increasing the bottom waveguide layer thickness or introducing impurity blocking layers in the AlN homoepitaxial buffer layer \cite{chaudhuri_high-conductivity_2021}. Further, a large fraction of the pump energy is lost to heat due to the generation of phonons due to hot electron-hole pairs. Hence, pumping the laser bars closer to the band edge should significantly reduce the threshold intensity. Finally, it is noted that the cavity loss for the laser bars discussed here are limited by the single semiconductor-air interface, resulting in the mirror loss of $\alpha_{m} \approx 14$ cm$^{-1}$ for a cavity length of 1 mm, in the ideal case. The threshold pump intensity can be further reduced by reducing this mirror loss. General techniques that are used are high-reflectivity oxides-based distributed Bragg reflectors enabled by atomic layer deposition \cite{sakai_-wafer_2020}, or distributed feedback lasers enabled by electron beam lithography \cite{muziol_distributed-feedback_2020}, but these techniques are mode-selective.

In summary, we achieved multi-mode lasing at $\approx$ 284 nm from side-emitting, optically injected, double heterostructure AlGaN laser structures. This is the first of its kind grown by plasma-assisted molecular beam epitaxy on bulk AlN. The importance of process optimization for Fabry-P\'erot lasers emitting at decreasing wavelengths is highlighted by a model for the combined effect of facet slant and roughness on the optical mirror loss. To this purpose, a combination of a dry etch or cleaving followed by a TMAH etch, is utilized, resulting in smooth and vertical facets and mesa sidewalls. Along with electronic design optimization to maximize optical gain, as well as waveguide design and growth optimization to keep the intrinsic optical losses low, the minimization of cavity losses is important in order to achieve room temperature, electrically injected, deep-ultraviolet, continuous-wave laser diodes based on the AlGaN semiconductor materials family.

The data that support the findings of this study are available from the corresponding author upon reasonable request.

The authors thank Professor Farhan Rana, Dr. Shyam Bharadwaj, Dr. Kazuki Nomoto, and Dr. Jimy Encomendero at Cornell University for helpful discussions. This work is partially supported by the Cornell Center for Materials Research with funding from the NSF MRSEC program (No. DMR-1719875). Further support is granted by ULTRA, an Energy Frontier Research Center funded by the U.S. Department of Energy (DOE), Office of Science, Basic Energy Sciences (BES), under Award No. DE-SC0021230. The authors also acknowledge the Cornell NanoScale Facility for device fabrication, supported by the National Science Foundation with grant No. NNCI-1542081 and RAISE-TAQS 1839196, as well as MRI 1631282.

\bibliography{main}

\end{document}